\documentclass[11pt]{article}
\usepackage[top=1in, left=1in, right=1in, bottom=1in]{geometry}
\usepackage{graphicx}
\usepackage{amssymb}
\usepackage{amsmath}

\title{Structured Random Binding: \\ a minimal model of protein-protein interactions}
\author{Ling-Nan Zou \\ Email: \texttt{zouln@psu.edu} \\ Department of Chemistry, The Pennsylvania State University \\ University Park, PA 16801. USA.}

\begin{document}

\maketitle

\begin{abstract}
We describe Structured Random Binding (SRB), a minimal model of protein-protein interactions rooted in the statistical physics of disordered systems. In this model, nonspecific binding is a generic consequence of the interaction between random proteins, exhibiting a phase transition from a high temperature state where nonspecific complexes are transient and lack well-defined interaction interfaces, to a low temperature state where the complex structure is frozen and a definite interaction interface is present. Numerically, weakly-bound nonspecific complexes can evolve into tightly-bound, highly specific complexes, but only if the structural correlation length along the peptide backbone is short; moreover, evolved tightly-bound homodimers favor the same interface structure that is predominant in real protein homodimers. 
\end{abstract}

\noindent In cells, protein-protein interactions are vital because proteins generally execute their activities in complex with other proteins \cite{noorenDiversityProteinproteinInteractions2003}. The textbook picture where proteins are depicted as rigid bodies interacting via interlocking surfaces, and where the specificity of an interaction originates from the specific geometry of its interface \cite{albertsMolecularBiologyCell2002}, collides uncomfortably with the fact pulldown experiments targeting one specific protein will routinely co-precipitate hundreds of additional proteins. How can any protein have sufficient surface structure to mediate this many interactions? Experimentally, detectable but weak interactions are often dismissed ad hoc as ``nonspecific'', but what distinguishes nonspecific interactions from weak but specific interactions? Here, we describe Structured Random Binding (SRB), a minimal model of protein-protein interactions where \textit{nonspecific} binding is a generic consequence of the interaction between \textit{random} proteins. SRB exhibits a phase transition from a high temperature state where nonspecific complexes are transient and lack well-defined interaction interfaces, to a low temperature state where the complex structure is frozen and a definite interface is present. In numerical simulations, weakly-bound nonspecific complexes can evolve into tightly-bound, highly specific complexes, but only if the structural correlation length along the peptide backbone is short. Furthermore, evolved tightly-bound homodimers favor the same interface structure that is predominant in real protein homodimers. Our model thus captures two salient features of protein-protein interactions --- the ubiquity of nonspecific complexes, and the structure of homodimers --- that cut across the diversity of protein species, and root these phenomena firmly in the statistical physics of disordered systems.

\section*{Defining the model}

\begin{figure}[t]
\begin{center}
\includegraphics{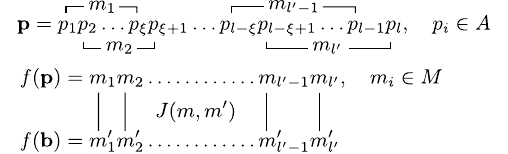}
\caption{\label{Fig1}Structured Random Binding model of protein-protein interactions. Sliding $\xi$-tuples of the primary sequence $\mathbf{p}$ maps onto structural motifs $m_i\in M$, and endows $\mathbf{p}$ with a secondary structure $f(\mathbf{p})$. Paired motif-motif interactions, encoded by $J(m, m')$, between two secondary structures (shown in parallel orientation) sum to yield the peptide-peptide binding energy $u(\mathbf{p}, \mathbf{b})$.}
\end{center}
\end{figure} 

To begin, we first consider the interaction between a small peptide and a protein. We represent peptide $\mathbf{p} = p_1 p_2 \dots p_l$ as a sequence of length $l$ (the primary sequence), where $p_i$ are letters drawn from a primary alphabet $A$ (the 20 proteinogenic amino acids). We represent protein $\mathbf{B}$ as a collection of $n_\mathbf{B}$ peptide epitopes (also of length $l$), $\mathbf{B} = \{\mathbf{b}^\alpha, \alpha=1,\dots, n_\mathbf{B}\}$, where $\mathbf{b}^\alpha = b_1^\alpha b_2^\alpha \dots b_l^\alpha$,  $b_i^\alpha \in A$; since only surface exposed residues participate in protein-protein interactions, $n_\mathbf{B}$ is proportional to the protein surface area. The interaction between residues on the peptide, and residues on the protein, depends on their respective structural contexts.  We define a map $f:A^\xi \to M$ from $\xi$-tuples of $A$ to structural motifs,
\begin{equation}
f(p_i p_{i+1} \dots p_{i+\xi-1}) = m_i,  p_i\in A,  m_i\in M,
\end{equation}
where $M$ is the motif alphabet, and $\xi$ is the structural correlation length along the peptide backbone; we assume $f$ is bijective and $\|M\| = \|A\|^\xi$. By acting on each sliding $\xi$-tuple of $\mathbf{p}$, $f$ endows $\mathbf{p}$ with a secondary structure $f(\mathbf{p}) = m_1 m_2\dots m_{l-\xi+1}$, $m_i\in M$ (Fig.~\ref{Fig1}). We define the binding energy $u(\mathbf{p}, \mathbf{b})$ between $\mathbf{p}$ and $\mathbf{b} \in \mathbf{B}$ as the sum of $l'=l-\xi+1$ paired motif-motif interactions,
\begin{align}
u_{\uparrow\!\uparrow}(\mathbf{p}, \mathbf{b})  &= \sum_{i=1}^{l'} J[f(\mathbf{p})_i, f(\mathbf{b})_i], \\
u_{\uparrow\!\downarrow}(\mathbf{p}, \mathbf{b})  &= \sum_{i=1}^{l'} J[f(\mathbf{p})_{l'-i+1}, f(\mathbf{b})_i],
\label{peptide_peptide_u}
\end{align}
depending on whether $\mathbf{p}$, $\mathbf{b}$ are oriented in parallel or antiparallel. Here, $J(m, m')$, $m, m' \in M$, is a $\| M\| \times \|M\|$ real symmetric matrix, whose elements are independently and identically distributed random variables with mean $\langle J\rangle = 0$, and finite variance $\sigma_J^2$ that sets the energy scale of motif-motif interactions. In the low temperature limit $T\to 0$, the binding energy $ U(\mathbf{P},\mathbf{B})$ of the binary complex $\mathbf{PB}$ is given by
\begin{equation}
U(\mathbf{P},\mathbf{B}) = \mathrm{min}\{u(\mathbf{p},\mathbf{b}), \mathbf{p} \in \mathbf{P}, \mathbf{b} \in \mathbf{B}  \}.
\label{protein_protein_U}
\end{equation}

\section*{Binding energy of nonspecific complexes}

\begin{figure}[t]
\begin{center}
\includegraphics{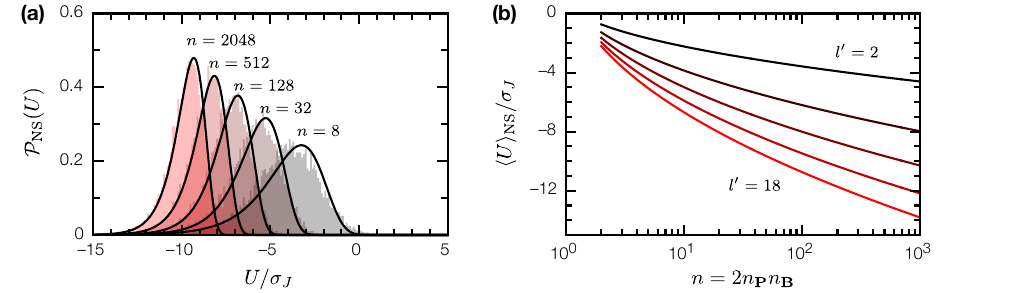}
\caption{\label{Fig2}The binding energy $U$ for nonspecific binary complexes of random proteins. (a) The distribution $\mathcal{P}_\mathrm{NS}(U)$ for nonspecific complexes; $n=2n_\mathbf{P}n_\mathbf{B}$ is twice the product of epitope counts for each protein. Filled histograms are empirical distributions from numerical simulations; solid lines are Gumbel distributions given by Eq.~\ref{protein_protein_U_distribution}, with $\mu$, $\phi$ given by Eqs.~\ref{location}, \ref{shape}. (b) Mean nonspecific binding energy $\langle U\rangle_\mathrm{NS}$ as a function of $n$, for epitope sizes $l'=2,6,10,14,18$.}
\end{center}
\end{figure} 

These ingredients are sufficient to fix the binding energy distribution $\mathcal{P}_\mathrm{NS}(U)$ for nonspecific binary complexes of random proteins. Since $u(\mathbf{p}, \mathbf{b})$ between a random peptide $\mathbf{p}$, and an epitope $\mathbf{b} \in \mathbf{B}$ on a random protein, is the sum of $l' = l -\xi +1$ random variables $J(m, m')$, for any choice of $\mathcal{P}(J)$ with zero mean and finite variance $\sigma_J^2$, in the limit $l' \to \infty$, $u$ will be normally distributed  
\begin{equation}
\mathcal{P}(u) = \frac{1}{\sqrt{2\pi l' \sigma_J^2}} \ \mathrm{exp}\left(-\frac{u^2}{2 l' \sigma_J^2}\right).
\label{peptide_peptide_u_distribution}
\end{equation}
For a binary complex of two random proteins $\mathbf{P}$, $\mathbf{B}$, its binding energy $U(\mathbf{P}, \mathbf{B})$ will follow the extreme value distribution for sample minima of $n = 2 n_\mathbf{P} n_\mathbf{B}$ point samples from $\mathcal{P}(u)$. In the limit $n\to \infty$, this is a Gumbel distribution \cite{colesIntroductionStatisticalModeling2001, kotzExtremeValueDistributions2015},
\begin{equation}
\mathcal{P}_\mathrm{NS}(U) = \frac{1}{\phi} \ \mathrm{exp}\left[-\frac{U-\mu}{\phi} + \mathrm{exp}\left(-\frac{U-\mu}{\phi}\right)\right],
\label{protein_protein_U_distribution}
\end{equation}
with location $\mu$ and scale $\phi$ parameters given by
\begin{equation}
\mu= -\sqrt{2l' \sigma_J^2} \  \mathrm{erf}^{-1}\left(1-2/n\right),
\label{location}
\end{equation}
\begin{equation}
\phi = -\mu - \sqrt{2l' \sigma_J^2} \ \mathrm{erf}^{-1}\left(1-2/ne\right).
\label{shape}
\end{equation}
The average nonspecific binding energy over all SRB realizations (``disorder averaged'') is
\begin{equation}
\langle U \rangle_\mathrm{NS} = \mu + \gamma \phi, \label{mean_nonspecific_binding_energy}
\end{equation}
where $\gamma$ is the Euler–Mascheroni constant. Note that neither $\mathcal{P}(u)$ nor $\mathcal{P}_\mathrm{NS}(U)$ depends on the size of the primary alphabet $\|A\|$ (assuming it is nontrivial). The weight in $\mathcal{P}_\mathrm{NS}(U)$ for $U > 0$ is negligible (Fig.~\ref{Fig2}a); the average binding energy $\langle U\rangle_\mathrm{NS}$ is strictly negative and decreases monotonically with increasing $n$ (Fig.~\ref{Fig2}b). Thus in the low-$T$ limit, nonspecific binding is a generic consequence of the interaction between random proteins. We can understand this intuitively as follows: given two random proteins, if their surfaces are large enough, we can always find a pair of epitopes, one on each protein, that will bind each other, even if the average epitope-epitope interaction is zero.

\section*{The condensation transition}

\begin{figure*}[t]
\begin{center}
\includegraphics{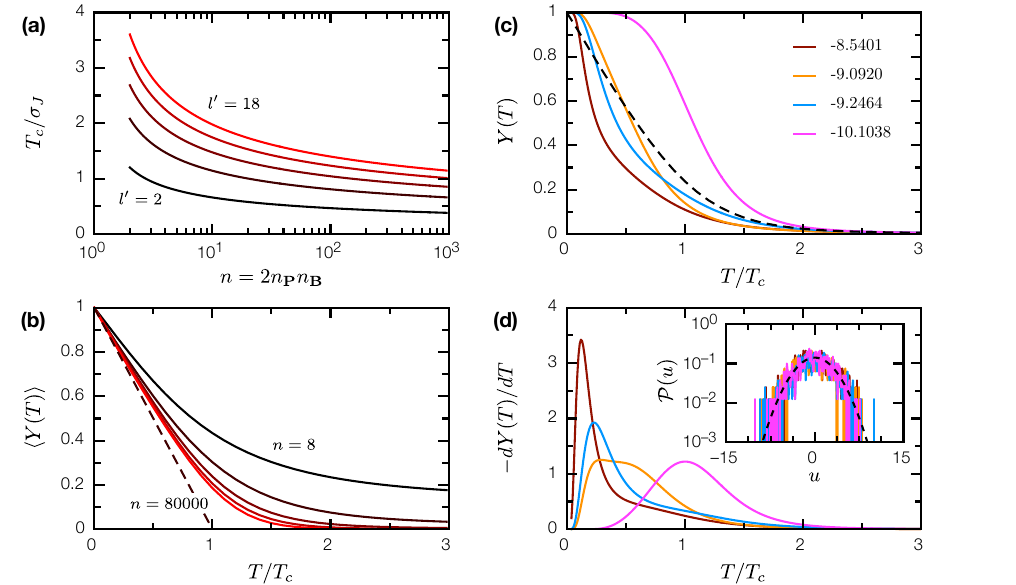}
\caption{\label{Fig3}The condensation transition in SRB. (a) Critical temperature $T_c$ as a function of $n=2n_\mathbf{P}n_\mathbf{B}$, for epitope sizes $l'=2,6,10,14,18$. (b) The mean participation ratio $\langle Y(T)\rangle$  as a function of $T/T_c$ for $n=8,72,800,7200,80000$. In the limit $n\to \infty$, $\langle Y(T)\rangle \to 1-T/T_c$ for $T<T_c$ (dashed line). (c) $Y(T)$ for 4 specific realization of random proteins $\mathbf{P}$, $\mathbf{B}$, and $J$ ($l = 10$, $\xi = 3$, and $n_\mathbf{P} = n_\mathbf{B}=20$); the binding energy $U(\mathbf{P},\mathbf{B})$ of the $\mathbf{PB}$ complex is indicated in the legend. Dashed line is the average $\langle Y(T)\rangle$ for $n=800$. (d) $-dY(T)/dT$ for the same realizations; the location of the peak is usually interpreted as the melting temperature of the protein complex. Inset: energy level distributions $\mathcal{P}(u)$ for these realizations; dashed line is the asymptotic distribution (Eq.~\ref{peptide_peptide_u_distribution}).}
\end{center}
\end{figure*} 

These results describes the  low-$T$ limit of SRB. To obtain finite $T$ behavior, we note that the interaction between random proteins $\mathbf{P}$, $\mathbf{B}$ describes a $n=2n_\mathbf{P}n_\mathbf{B}$ level system whose energy levels are independently and identically distributed per $\mathcal{P}(u)$, with density of states given by
\begin{equation}
\langle\rho(u)\rangle = n\mathcal{P}(u) = \frac{e^{\mathrm{log}(n) - \frac{u^2}{2l'\sigma_J^2}}}{\sqrt{2l'\sigma_J^2}}.
\end{equation}
So stated, the equilibrium statistical physics of SRB is analogous to that of the exactly solvable Random Energy Model (REM) \cite{derridaRandomEnergyModelLimit1980, derridaRandomenergyModelExactly1981}.  Originally formulated as a minimal model of spin glasses, REM has also proven influential in informing the statistical physics of protein folding \cite{shakhnovichImplicationsThermodynamicsProtein1990, onuchicTheoryProteinFolding1997}. In the thermodynamic limit, REM undergoes a phase transition, known as condensation, at temperatures $T<T_c$ to a frozen phase where the Boltzmann measure condenses onto a smaller-than-exponential set of configurations. In SRB, the corresponding $T_c$ (in the limit $n\to \infty$) is
\begin{equation}
\label{Tc}
T_c = \sigma_J\sqrt{\frac{l' }{2\ \mathrm{log}(n)}},
\end{equation} 
and is slowly varying in both $n$ and $l'$ (Fig.~\ref{Fig3}a). For $T>T_c$, the complex between two random proteins is dynamic and lacks a definite interaction interface; for $T<T_c$, the complex freezes into a fixed conformation and a definite interface is present. The participation ratio
\begin{equation}
Y(T) = \frac{\sum_{\{\mathbf{p} \in \mathbf{P}, \mathbf{b} \in \mathbf{B}\}}  e^{-2u(\mathbf{p}, \mathbf{b})/T}}{\left(\sum_{\{\mathbf{p} \in \mathbf{P}, \mathbf{b} \in \mathbf{B}\}} e^{-u(\mathbf{p}, \mathbf{b})/T}\right)^2}
\end{equation}
is the probability that two samples of the system return the same configuration; thus $1/Y(T)$ is effectively the number of states that dominate the Boltzmann measure. In REM, $\langle Y(T)\rangle = 0$ for $T>T_c$, $\langle Y(T)\rangle = 1-T/T_c$ for $T<T_c$ \cite{derridaSampleSampleFluctuations1985, mezardRandomFreeEnergies1985}; SRB exhibits the same behavior as $n\to\infty$ (Fig.~\ref{Fig3}b).

For SRB realizations with finite-sized proteins $\mathbf{P}$, $\mathbf{B}$, the  condensation phase transition broadens into crossover representing the binding/unbinding of the $\mathbf{PB}$ complex. Remarkably, individual SRB realizations, founded on the same underlying $\mathcal{P}(J)$, can have drastically different $Y(T)$ curves and apparent melting temperatures (where $-dY/dT$ is maximal); only upon averaging do we recover the REM-like transition (Figs.~\ref{Fig3}c, d). This chaotic behavior is a direct consequence of the condensation transition, where the $T<T_c$ thermodynamics of SRB is dominated by a small number of low-$u$ configurations. Different SRB realizations, despite their statistically similar energy spectra, have distinct thermodynamics that arise entirely from subtle differences in the low-$u$ tail (Fig.~\ref{Fig3}d, inset).

\section*{Evolution of tightly-bound complexes}

In SRB, nonspecific complexes have a well-defined energy scale $\langle U\rangle_\mathrm{NS}$. How can we obtain tightly-bound complexes with $U \ll \langle U\rangle_\mathrm{NS}$? Computationally, this is an optimization problem: given $J$ and peptide $\mathbf{b}$, find peptide $\mathbf{p}$ such that $u(\mathbf{p}, \mathbf{b})$ is minimal. In the limit $\xi \to 1$, single residue changes in $\mathbf{p}$ results in single motif changes in $f(\mathbf{p})$, and $u(\mathbf{p}, \mathbf{b})$ can be minimized term by term. In the opposite limit $\xi \to l$, single residue changes in $\mathbf{p}$ completely alters $f(\mathbf{p})$, and there can be no correlation between binding energy and primary sequence; minimizing $u(\mathbf{p}, \mathbf{b})$ requires exhaustive search through all possible primary sequences. In the regime $1<\xi<l$, fixing $f(\mathbf{p})$ at motif $f(\mathbf{p})_i$, by fixing $\mathbf{p}$ at residues $p_i$, ..., $p_{i+\xi-1}$, also restricts the possible values of motifs $f(\mathbf{p})_{i-\xi+1}$, ..., $f(\mathbf{p})_{i+\xi-1}$. This means while every primary sequence $\mathbf{p}$ can be mapped to a secondary structure $f(\mathbf{p})$, arbitrary motif sequences $\mathbf{m} = m_1 m_2\dots m_{l'}$, $m_i \in M$, are generally noninvertible, i.e. $\mathbf{p} = f^{-1}(\mathbf{m})$ does not exist. Minimizing $u$, which is a function of motif sequences, must be restricted to invertible secondary structures, which forms a small (fraction = $\|A\|^{l-(l-\xi+1)\xi}$) and sparse (single motif change on an invertible secondary structure renders it noninvertible) subset in the space of motif sequences. As is often the case with constrained optimization \cite{mezardInformationPhysicsComputation2009}, we conjecture there is no fast algorithm for solving SRB in this regime.

\begin{figure}[t]
\begin{center}
\includegraphics{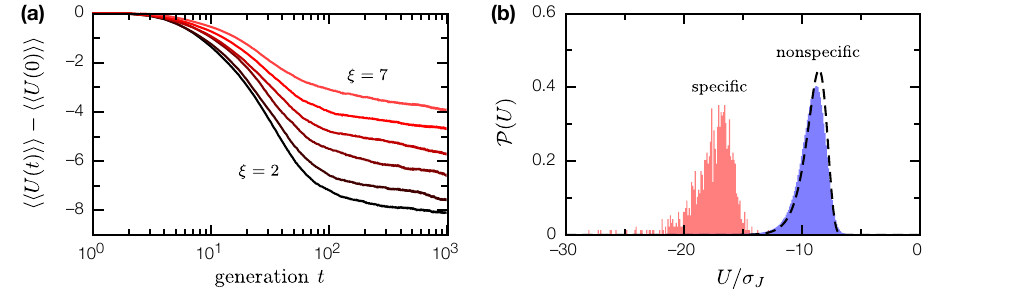}
\caption{\label{Fig4}Evolution of weakly-bound, nonspecific complexes of random proteins into tightly-bound, highly specific complexes. (a) Binding energy evolution of the nonspecifc complex between random proteins $\mathbf{P}$, $\mathbf{B}$ ($l=10$, $n_\mathbf{P} = n_\mathbf{P} =20$) over $10^3$ generations, for $\xi = 2,3,4,5,6,7$; $\langle\langle U\rangle\rangle$ is doubly averaged over the ensemble of $N=10^3$ binary complex mutants, and over $10^2$ SRB realizations. (b) Distribution of binding energies $\mathcal{P}(U)$ for coevolved proteins $\mathbf{P}$, $\mathbf{B}$ ($\xi = 3$), in binary complex with each other (pink fill), and in binary complex with other random proteins (blue fill). Dashed line is the binding energy distribution $\mathcal{P}_\mathrm{NS}(U)$ for nonspecific complexes of random proteins.}
\end{center}
\end{figure} 

Biologically, tightly-bound protein complexes are products of evolution. We simulated affinity evolution in a binary complex of two coevolving random proteins $\mathbf{P}$, $\mathbf{B}$; for tractable numerics, we restrict ourselves to a small primary alphabet $\|A\| = 4$. We start from an uniform ensemble $\{\mathbf{P}^\alpha\mathbf{B}^\alpha, \alpha = 1, \dots, N\}$, $\mathbf{P}^\alpha\mathbf{B}^\alpha = \mathbf{P}^\beta\mathbf{B}^\beta$ for all $\alpha$, $\beta$. At each generation $t$, we mutate the binary complex (at rate $\epsilon = 0.01$ per residue per generation), and apply selection with an affinity-dependent survival probability
\begin{equation}
\Theta(U; U_0) = \frac{1}{e^{(U - U_0)/\sigma_J} + 1}.
\end{equation}
Here, the selection threshold $U_0(t) = \langle U[\mathbf{P}(t), \mathbf{B}(t)]\rangle$ coevolves with the binary complex ensemble to maintain selection pressure. As shown in Fig.~\ref{Fig4}a, the long time evolutionary dynamics is characterized by a slow $\mathrm{log}(t)$ behavior similar to aging dynamics in glassy systems \cite{berthierTheoreticalPerspectiveGlass2011}; while steady state could not be reached within the simulation window, it is clear that evolution is most effective in enhancing binding affinity when the structural correlation length $\xi$ is short. Where $\mathbf{PB}$ evolved into tightly-bound complexes, the resulting interactions are highly specific: while the coevolved proteins bind each other with $U\ll \langle U\rangle_\mathrm{NS}$, their binding energies with other random proteins is indistinguishable from that between any two random proteins (Fig.~\ref{Fig4}b). This result suggests we can make a physically principled distinction between specific and nonspecific protein complexes: specific complexes are those whose binding energy $U$ is unlikely to be the result of random protein-protein interactions. It also suggests the evolution of tightly-bound, highly specific protein complexes is not ex nihlio, but emerges from a basis of ubiquitous nonspecific interactions.

\section*{Structural bias in homodimeric complexes}

\begin{figure}[t]
\begin{center}
\includegraphics{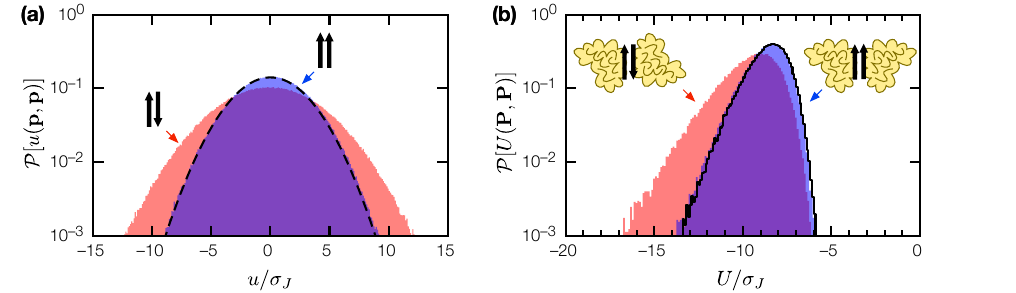}
\caption{\label{Fig5}Structural bias in homodimeric complexes. (a) Binding energy distribution $\mathcal{P}[u(\mathbf{p}, \mathbf{p})]$ for antiparallel (pink fill) and parallel (blue fill) isologus peptide-peptide interactions, for random peptides ($l=10$, $\xi=3$). Dashed line indicates $\mathcal{P}(u)$ for heterlogous peptide-peptide interactions. (b) Empirical binding energy distributions $\mathcal{P}(U)$ for random homodimers $\mathbf{PP}$ ($l=10$, $\xi=3$, $n_\mathbf{P} = 20$) with antiparallel isologus (pink fill), parallel isologus (blue fill), and heterlogus (black line) dimer interfaces.}
\end{center}
\end{figure}

Compared to heterodimeric interactions, homodimeric interactions in SRB exhibit a subtle but significant distinction. For homodimers $\mathbf{PP}$ with dimer interface $\mathbf{p}, \mathbf{p}' \in \mathbf{P}$, if $\mathbf{p} \ne \mathbf{p}'$ (heterologus), or if $\mathbf{p} = \mathbf{p}'$ (isologus) are parallel, then $u(\mathbf{p}, \mathbf{p}')$ and $u_{\uparrow\!\uparrow}(\mathbf{p}, \mathbf{p})$ will be the sum of $l'$ random variables $J(m, m')$. However, when $\mathbf{p} = \mathbf{p}'$ are antiparallel, $u_{\uparrow\!\downarrow}(\mathbf{p}, \mathbf{p})$ is twice the sum of $l'/2$ random variables. From the arithmetic of random variables, $\mathcal{P}[u_{\uparrow\!\downarrow}(\mathbf{p}, \mathbf{p})]$ will have twice the variance of $\mathcal{P}[u_{\uparrow\!\uparrow}(\mathbf{p}, \mathbf{p})]$ or $\mathcal{P}[u(\mathbf{p}, \mathbf{p}')]$, $\mathbf{p} \ne \mathbf{p}'$ (Fig.~\ref{Fig5}a). This means random homodimers with antiparallel isologus interfaces are on average more tightly-bound than homodimers in other configurations (Fig.~\ref{Fig5}b). As weakly-bound nonspecific homodimers evolve into tightly-bound homodimers, they tend to retain antiparallel isologus interfaces (if present in the ancestral complex), or adopt it de novo, such that antiparallel isologus interfaces are predominant in tightly-bound homodimers. For $l'=8$, $\xi = 3$, $n_\mathbf{P} = 20$, random homodimers as generated were 33\% antiparallel isologus, 3.3\% parallel isologus, and 63.7\% heterologus; after 1000 generations of affinity evolution, the resulting tightly-bound homodimers are 98\% antiparallel isologus, 1\% parallel isologus, and 1\% heterologus. Real protein homodimers predominantly have antiparallel isologus interfaces \cite{goodsellStructuralSymmetryProtein2000, bahadurDissectingSubunitInterfaces2003}; our results supports the view that this phenomenon results from an intrinsic bias in homodimeric protein-protein interactions \cite{lukatskyStatisticallyEnhancedSelfAttraction2006, lukatskyStructuralSimilarityEnhances2007, andreEmergenceSymmetryHomooligomeric2008}. In the case of SRB, this bias is entirely statistical in origin.

\section*{Discussion}

SRB is a vastly simplified model of protein-protein interactions. Actual proteins are often somewhat conformationally flexible, and can undergo conformational changes upon complexation \cite{keskinPrinciplesProteinProteinInteractions2008}; this has entropic effects which SRB neglects. Many proteins have interaction surfaces made up of physically proximal, but sequentially distal residues, and the structural context of a surface-exposed residue depends on more than just nearby residues along the peptide backbone. Nevertheless, there is evidence the core physical insight of SRB, that the interaction between two proteins describes a multi-level system with a random or weakly-correlated energy spectrum, holds for real protein complexes \cite{janinQuantifyingBiologicalSpecificity1996, bernauerDockingAnalysisStatistical2005}. If SRB does indeed capture the statistical physics of real protein-protein interactions (at least qualitatively), several implications are immediately evident:

Suppose the energy scale $\sigma_J \sim k_B \times 300\mathrm{K}$, then at physiological temperature, a significant fraction proteins inside a cell will be bound into nonspecific complexes that have no functional significance. If true, then we must reassess the capabilities and limitations of protein pulldown, proximity labelling, and two-hybrid experiments in identifying functional protein complexes \cite{gingrasAnalysisProteinComplexes2007, qinDecipheringMolecularInteractions2021, suterTwoHybridTechnologies2012}. While nonspecific protein complexes will be weakly-bound, they are ubiquitous due to the generic nature of nonspecific binding, and are likely to dominate any experimental signal. Here, the physically principled distinction between specific and nonspecific complexes offered by SRB can inform new experiments and analyses seeking to identify functional protein-protein interactions \cite{heMultiplexMappingProteinprotein2025}. 

Proteins in the cytoplasm, and in dense protein solutions, undergo anomalous diffusion where $\langle \Delta r^2\rangle \propto t^\beta$, $\beta < 1$, and their rotational diffusion are more impeded in dense protein solutions than in simple solvents of similar viscosity \cite{wangEffectsProteinsProtein2010, roosCouplingDecouplingRotational2016, hoflingAnomalousTransportCrowded2013, grimaldoDynamicsProteinsSolution2019}. SRB suggests we should not expect the free diffusion of proteins in dense protein solutions, as they will be mostly bound into nonspecific complexes. This will have an especially pronounced effect for rotational diffusion: while the translational diffusion constant $D_t \propto a^{-1}$, where $a$ is the hydrodynamics radius, the rotational diffusion constant $D_r \propto a^{-3}$. Additionally, due to the condensation transition, anomalous transport in dense protein solutions should be even more pronounced at low temperatures, above and beyond effects expected from temperature-dependent changes in solvent viscosity.

Finally, the sensitivity of protein complex thermodynamics to the exact disposition of low energy states suggests the accurate prediction of protein complex binding free energies, which is a long standing challenge in computational biology \cite{kastritisAreScoringFunctions2010, siebenmorgenComputationalPredictionProtein2020}, may be fundamentally limited. Here, it is possible that different choices of necessary approximations, e.g. for molecular geometry, force fields, etc., leading to subtle differences in low lying states, can nevertheless result in large differences in computed thermodynamic properties.

Molecular investigations of protein complexes are singularly focused on the individuality of the proteins involved. Our work demonstrates that statistical physics, encapsulated in a minimal model, can illuminate features of protein-protein interactions that cut across the diversity of protein species. SRB suggests nonspecific protein-protein binding is an ineluctable physical phenomenon, and that the proteins inside a cell are ever participants in a plethora of specific and nonspecific complexes. While these ubiquitous interactions challenge scientists who seek to decipher specific functions of detected interactions, they may be biologically significant in a broader sense. We have already suggested nonspecific interactions could be the source from which highly specific complexes evolved; we speculate nonspecific interactions can also be a mechanism that cells can exploit, to enable subtle modes of subcellular organization based on physical principles.

\paragraph{Acknowledgements.} We are grateful to Stephen Benkovic for his advice and support. We also benefited from discussions with Tom Witten and Nathan Keim. This work was supported by the U.S. National Institutes of Health under award R01GM024129.

\bibliographystyle{unsrt}
\bibliography{SRB}

\end{document}